\documentclass[11pt]{article}
\title{The Interaction of Dirac Particles with Non-Abelian Gauge Fields
and Gravity -- Black Holes}
\author{Felix Finster, Joel Smoller\thanks{Research supported in part by the
NSF, Grant No.\ DMS-G-9802370.}, and Shing-Tung
Yau\thanks{Research supported in part 
by the NSF, Grant No.\ 33-585-7510-2-30.}}
\date{October 1999}

\newtheorem{Def}{Def.}[section]
\newtheorem{Thm}[Def]{Theorem}
\newtheorem{Prp}[Def]{Proposition}
\newtheorem{Lemma}[Def]{Lemma}

\newtheorem{Corollary}[Def]{Corollary}
\newcommand{\Proof}{{\em{Proof: }}}
\newcommand{\QED}{\ \hfill $\FBox$ \\[1em]}

\newcommand{\spc}{\;\;\;\;\;\;\;\;\;\;}
\newcommand{\bra}{\mbox{$< \!\!$ \nolinebreak}}
\newcommand{\ket}{\mbox{\nolinebreak $>$}}

\newcommand{\FBox}{\rule{2mm}{2.25mm}}

\def\rdots{\mathinner{\mkern1mu\raise1pt\vbox{\kern1pt\hbox{.}}\mkern2mu
   \raise4pt\hbox{.}\mkern2mu\raise7pt\hbox{.}\mkern1mu}}
\newcommand{\ve}{\varepsilon}

\newcommand{\al}{\alpha}
\newcommand{\om}{\omega}
\newcommand{\ra}{\rightarrow}
\newcommand{\Z}{{\rm Z\kern-.35em Z}}
\newcommand{\bP}{{\rm I\kern-.15em P}}
\newcommand{\Q}{\kern.3em\rule{.07em}{.65em}\kern-.3em{\rm Q}}
\newcommand{\R}{{\rm I\kern-.15em R}}
\newcommand{\h}{{\rm I\kern-.15em H}}
\newcommand{\C}{\kern.3em\rule{.07em}{.55em}\kern-.3em{\rm C}}
\newcommand{\T}{{\rm T\kern-.35em T}}

\begin{document}
\maketitle

\begin{abstract}
We consider a static, spherically symmetric system of a Dirac particle
in a classical gravitational and $SU(2)$ Yang-Mills field.  We prove
that the only black-hole solutions of the corresponding
Einstein-Dirac-Yang/Mills equations are the Bartnik-McKinnon black-hole
solutions of the $SU(2)$ Einstein-Yang/Mills equations; thus
the spinors must vanish identically. This indicates that the
Dirac particles must either disappear into the black-hole or escape to infinity. 
\end{abstract}

\section{Introduction}
Recently, the Einstein-Dirac-Yang/Mills (EDYM) equations were derived 
for a static, spherically symmetric system of a Dirac particle which 
interacts with both a gravitational field and the magnetic component 
of an $SU(2)$ Yang-Mills field \cite{8}. This system was constructed by 
choosing a representation of the rotation group which acts 
non-trivially on the YM index, and by making an ansatz involving two real 
functions for the Dirac wave function, which is invariant under this group
representation. The resulting equations were shown
to admit stable particle-like solutions for physically relevant values
of the coupling constants. In this paper, we study black-hole solutions of
these EDYM equations.

We prove that the only black-hole solutions of our EDYM equations are
the Bartnik-McKinnon (BM) black holes; that is, the spinors must vanish
identically. In other words, the EDYM equations do not admit 
normalizable black-hole solutions. Thus in the presence of quantum mechanical
Dirac particles, static and spherically symmetric black-hole solutions do not
exist. Another interpretation
of our result is that Dirac particles can only either disappear
into the black-hole or escape to infinity.  These results are proved under
very weak regularity assumptions on the form of the event horizon; see
Assumptions (I)-(III) in the next section.

Our work here is a continuation of \cite{6} where we showed that the
Einstein-Dirac-Maxwell (EDM)
equations do not admit normalizable black hole solutions.
However, in contrast to the EDM system \cite{6}, where an electric field is 
present, we consider here the influence of a magnetic field, more 
precisely the influence of the magnetic component of a non-abelian gauge field.
Furthermore, we point out that the Dirac particle considered here has zero
total angular momentum. In analogy 
to \cite{6}, one could also form a spherically symmetric system out of 
$(2j+1)$ Dirac particles each having angular momentum $j$, 
$j=1,2,\ldots$. However, the EDYM equations for such a system would
involve four real spinor functions. This is a considerably more 
difficult problem, which we are presently investigating.

\section{The Coupled EDYM Equations.}
\setcounter{equation}{0}
\label{sec2}
In \cite{8}, we derived the spherically symmetric, static EDYM system of 
a Dirac particle and an $SU(2)$ YM field with vanishing electric 
component. We shall not repeat this derivation here, but merely 
write down the EDYM equations. We consider a Lorentzian 
metric in polar coordinates $(t,r, \vartheta, \varphi)$ of the form
\[ ds^2 = \frac{1}{T(r)^2}\: dt^2 \:-\: \frac{1}{A(r)}\:dr^2
\:-\: r^2 (d\vartheta^2 + sin^2\vartheta \:d\varphi^2) \]
with positive functions $A$ and $T$.
The Dirac wave function is described by two real functions
$(\alpha(r), \beta(r))$, and the potential $w(r)$ corresponds to the magnetic 
component of an $SU(2)$ YM field. Then the EDYM equations are
\begin{eqnarray}
\sqrt{A} \:\alpha^\prime &=& \frac{w}{r} \:\alpha \:-\: (m + \omega T) \:
\beta \label{2.1} \\
\sqrt{A} \:\beta^\prime &=& (-m+\omega T) \:\alpha \:-\: \frac{w}{r} \:\beta
\label{2.2} \\
r \:{A^{\prime }}
&=& 1-A \:-\:\frac{1}{e^2}\:\frac{{{(1-{w^2})}^2}}{{r^2}}
\:-\:2 \omega  {T^2} \big({{\alpha }^2}+ {{\beta }^2}\big)
\:-\:\frac{2}{e^2} \:A w'^2  \label{2.3} \\
2 r A \:\frac{ {T^{\prime }}}{T} &=& -1+A \:+\:
\frac{1}{e^2} \:\frac{{{(1-{w^2})}^2}}
{{r^2}} \:+\:2 m T \big({{\alpha }^2}-{{\beta }^2}\big)
\:-\:2 \omega  \:{T^2} \big({{\alpha }^2}+{{\beta }^2}\big)
\nonumber \\
&&+4 \:\frac{T}{r} \:w\:\alpha \beta
\:-\:\frac{2}{e^2} \:A w'^2 \label{2.4} \\
{r^2} A \:{w^{\prime \prime}} &=& -\big(1-{w^2}\big) \:w\:+\:e^2\:r T \alpha
\beta \:-\: {r^2} \:\frac{A^\prime \:T - 2 A \:T^\prime}{2T} \:w^\prime \;\;\;.
\label{2.5}
\end{eqnarray}
Equations (\ref{2.1}) and (\ref{2.2}) are the Dirac equations, (\ref{2.3}) and
(\ref{2.4}) are the Einstein equations, and (\ref{2.5}) is the Yang-Mills
equation.  Here the constants $m$, $\omega$, and $e$ are the rest mass of
the Dirac particle, its energy, and the YM coupling constant, respectively.

We shall here consider black-hole solutions of this EDYM system. Thus 
we assume as in \cite{6} that the surface $r = \rho > 0$ is a black-hole event
horizon,
\begin{equation}
A(\rho) = 0 \spc {\rm and} \spc A(r) > 0 \ \ \ {\rm if} \
\ \ r > \rho \;\;\; . \label{2.6}
\end{equation}
In addition, we assume the following conditions:
\begin{equation}
\int^\infty_{r_0} (\al^2 + \beta^2) \:\frac{\sqrt{T}}{A} \:dr
\;<\; \infty \spc {\mbox{for every $r_0 > \rho$}}
\label{2.7}
\end{equation}
(the spinors are normalizable),
\begin{equation}
\lim_{r\ra \infty} \frac{r}{2} \:(1 - A(r)) \;<\; \infty
\label{2.8}
\end{equation}
(finite ADM mass),
\[ \lim_{r\ra \infty} T(r) \;=\; 1 \]
(the gravitational field is asymptotically flat Minkowskian),
\begin{equation}
\lim_{r\ra \infty} \big( w(r), w'(r) \big) \spc {\mbox{is finite}}
\label{2.10}
\end{equation}
(well-behavedness of the Yang-Mills field).

Concerning the event horizon $r=\rho$, we make the following three
regularity assumptions (cf.\ \cite{6}):
\begin{description}
\item[(I)] The volume element $\sqrt{|\det g_{ij}}| = |\sin \vartheta| \:
r^2 A^{-1} \:T^{-2}$ is smooth and non-zero on the horizon; i.e.
\[ T^{-2} A^{-1}, \ T^2A \in C^1 ([\rho, \infty)) \;\;\;. \]
\item[(II)] The strength of the Yang-Mills field $F_{ij} $ is given by
(cf.\ \cite{2})
\[ {\mbox{Tr}} (F_{ij} F^{ij}) \;=\; \frac{2A w'^2}{r^2}
+ \frac{(1-w^2)^2}{r^4} \;\;\;. \]
We assume that this scalar is bounded near the horizon; i.e. we 
assume that outside the event horizon and near $r=\rho$,
\begin{equation}
{\mbox{$w$ and $A w'^2$ are \ bounded.}} \label{2.12}
\end{equation}
\item[(III)] The function $A(r)$ is monotone increasing outside of and
near the event horizon.
\end{description}
As discussed in \cite{6}, if assumptions (I) or (II) were violated, an
observer freely falling into the black hole would feel strong forces
when crossing the horizon.  Assumption (III) is considerably weaker than
the corresponding assumption in \cite{6}; indeed, in
\cite{6} we assumed that the function $A(r)$ obeyed a power law $A(r) =
c\:(r-\rho)^s + {\cal{O}}((r-\rho)^{s+1})$, with positive constants $c$ and
$s$, for $r > \rho$.

\section{Non-Existence of Black-Hole Solutions.}
\setcounter{equation}{0}
Our main result is the following theorem.
\begin{Thm}
Every black-hole solution of the EDYM equations (\ref{2.1})--(\ref{2.5})
satisfying the regularity conditions (I), (II), and (III)
cannot be normalized, and coincides with a BM
black-hole of the corresponding Einstein-Yang/Mills (EYM) equations;
that is, the spinors $\al$ and $\beta$ must vanish identically
outside the event horizon.
\end{Thm}
We first make a few remarks:
\begin{description}
\item[(A)]  In \cite{12}, it was proved that any black-hole solution of
the EYM equations which has finite (ADM) mass (i.e., satisfies (\ref{2.8})),
must be one of the BM black-hole solutions whose existence
was first demonstrated in \cite{10}.  Thus amending the EYM equations by
taking quantum mechanical effects into account, in the sense that we allow
both the gravitational and Yang/Mills fields to interact with Dirac particles,
does not yield any new types of black-hole solutions.  
\item[(B)]  In \cite{6}, we studied black-hole solutions of the
Einstein-Dirac-Maxwell (EDM) equations, and we proved a corresponding
result for this (EDM) system.  As mentioned above, we assumed there that
near the event horizon, $A(r) = c \:(r-\rho)^s + {\cal{O}}((r-\rho)^{s+1})$,
where $c > 0$ and $s > 0$.  We were able to prove that if
$s \not= 2$, the spinors must vanish outside of the event horizon, but
if $s=2$, we could only obtain this result numerically.
Theorem 3.1 thus gives an even stronger result in the case of a YM 
field.
\end{description}
In the remainder of this paper, we shall give the proof of Theorem 3.1.
We assume that we have a black-hole
solution of the EDYM equations (\ref{2.1})--(\ref{2.10}), with event horizon at
$r=\rho$, satisfying the regularity assumptions (I)--(III), where the
spinors $(\al(r), \beta(r))$ do not vanish identically in the region
$r > \rho$.  We will show that this leads to a
contradiction.  We consider two cases:  either $A^{-\frac 1 2}$ is
integrable near the horizon, or $A^{-\frac 1 2}$ fails to be integrable
near the horizon.  We begin with\\[1em]
\underline{Case 1}.   $A^{-\frac 1 2}$ is integrable near $r=\rho$. \\[.5em]
In order to obtain the desired contradiction, we shall need a few
preliminary results, the first of which is 

\begin{Lemma}
If $A^{-\frac{1}{2}}$ is integrable near $r=\rho$, then
there exist positive constants $c, \ve$ such that
\begin{equation}
c \;\le\; \al^2(r) + \beta^2(r) \;\le\; \frac 1 c \spc {\mbox{if $\rho
< r < \rho + \ve$}} \;\;\;. \label{3.1}
\end{equation}
\end{Lemma}
{\Proof}
If we multiply (\ref{2.1}) by $\al$ and (\ref{2.2}) by $\beta$ and add, we
obtain
\begin{eqnarray}
\sqrt{A} \ \frac d {dr} (\al^2 + \beta^2) &=& 2
\left( \begin{array}{c} \alpha \\ \beta \end{array} \right)^T
\left( \begin{array}{cc}
\displaystyle \frac w r & -m \\
-m & \displaystyle -\frac{w} r
\end{array} \right)
\left( \begin{array}{c} \alpha \\ \beta \end{array} \right) \label{3.2} \\
&\le& 2 \:\sqrt{m^2 + \frac{w^2}{r^2}} \: (\al^2 + \beta^2) \;\;\;,
\nonumber
\end{eqnarray}
where in the last estimate we have computed the eigenvalues of the above
matrix. Since $(\al, \beta)$ is a non-trivial solution, the uniqueness theorem
for ODEs implies that $(\al^2 + \beta^2)(r)\not= 0$ on all intervals of the
form $(\rho, \rho+\ve)$. Thus dividing (\ref{3.2}) by $\sqrt{A}
\:(\al^2 + \beta^2)$ and integrating from $r > \rho$ to $\rho +\ve$ gives
\begin{equation}
|\log (\al^2 + \beta^2)(\rho+\ve) - \log(\al^2 + \beta^2)(r)| \;\le\; 2
\int^{\rho +\ve}_r A^{-\frac 1 2}(s) \:\sqrt{m^2 + \frac{w^2(s)}{s^2}
} \: ds \;\;\;.
\label{3.3}
\end{equation}
Since $w(r)$ is bounded near the event horizon (by Assumption (II)), and
$A^{-\frac 1 2}$ is integrable near the event horizon, we can take the 
limit $r \searrow \rho$ in (\ref{3.3}) to get the desired result.
\QED

\begin{Corollary}
If  $A^{-\frac 1 2}$ is integrable near $r=\rho$, then $\om=0$.
\end{Corollary}
{\Proof}
Combining (\ref{2.3}) and (\ref{2.4}), we have
\begin{equation}
r\: (AT^2)' = -4\om \:T^4 \:(\al^2+\beta^2) + T^3 \left[ 2m \:(\al^2-\beta^2) +
\frac{4w}{r} \:\al \beta \right] - \frac{4}{e^2} \: (A w'^2) \: T^2
\;\;\;. \label{3.4}
\end{equation}
From Assumption (II), together with the last result, we see
that the coefficients of $T^4, T^3$, and $T^2$ on the right-hand-side of
(\ref{3.4}) are bounded.  From Assumption (I) we see that the left-hand side
of (\ref{3.4}) is bounded near the event horizon. 
 Since $T(r) \ra \infty $ as $r \searrow \rho$ (again by (I)), we
conclude from (\ref{3.4}) and Lemma 3.2 that $\om = 0$.
\QED
In view of this last result, we can write the Dirac equations
(\ref{2.1}) and (\ref{2.2}) as
\begin{eqnarray}
\sqrt{A} \; \al' &=& \frac {w} r \:\alpha- m \:\beta \label{3.5} \\
\sqrt{A} \; \beta' &=& -m \:\al - \frac {w} r \:\beta\;\;\;. \label{3.6}
\end{eqnarray}  
If we multiply (\ref{3.5}) by $\beta$ and (\ref{3.6}) by $\al$, and add, we
obtain
\begin{equation}
\sqrt{A} \; (\al \beta)' \;=\; -m \:(\al^2+\beta^2) \;<\; 0 \;\;\;,
\label{3.7}
\end{equation}
so that $\al \beta$ is monotone decreasing. Thus $(\alpha \beta)(r)$ 
has a limit for $r \rightarrow \infty$ (which might be $-\infty$). 
Since $\alpha^2 + \beta^2 \geq 2 \:|\alpha \beta|$, we see that the 
normalization condition (\ref{2.7}) will be satisfied only if this 
limit is zero. We thus have proved the following lemma.
\begin{Lemma} If $\om = 0$, then $(\al \beta)(r)$ is a
positive decreasing function tending to $0$ as $r \ra \infty$.
\end{Lemma}
\underline{Remark}.  We did not use the fact that $A^{-\frac 1 2}$ is
integrable to prove this lemma. \\

The YM equation (\ref{2.5}) can be written equivalently as an equation
for $Aw'$,
\begin{equation}
r^2(Aw')' = -w \:(1-w^2) \:+\: e^2\:\frac{r(\sqrt{A}\; T)\al \beta}{\sqrt{A}}
\:+\: r^2 \:\frac {(AT^2)'}{2 AT^2} (Aw') \;\;\;.
\label{3.8}
\end{equation}
Since $A w'^2$ is bounded, we see that $A^2 w'^2 \ra 0$ as $r \searrow
\rho$ and thus $Aw' \ra 0$ as $r \searrow \rho$.  In view of the last
lemma, together with Assumption (I), we see that for $r$ near $\rho$, we
can write (\ref{3.8}) in the form
\begin{equation}
    (Aw')^\prime \;=\; d(r) \:+\: \frac {c(r)} {\sqrt{A(r) }}
\label{3.9}
\end{equation}
where $d(r)$ is bounded, and $c(r)$ is a positive function which is
bounded away from zero near the event horizon.  It thus follows from
(\ref{3.9}) that we have
\begin{equation}	
(Aw')' (r) \;\ge\; d + \frac c {\sqrt{A(r)}} \;\;\; ,
\label{3.10}
\end{equation}
for $r > \rho$ ($r$ near $\rho)$, where $d$ and $c$ are constants, and $c > 0$.
In order to analyze this inequality, we shall need the following lemma.
\begin{Lemma}
If $A(r)^{- \frac 1 2}$ is integrable near $r=\rho$, then there exists
a function $g(r)$ with $0<\lim_{r \searrow \rho} g(r) < \infty$, such 
that for $r$ near $\rho$,
\begin{equation} A(r) \;=\; (r-\rho) \:g(r) \;\;\;. \label{3.11}
\end{equation}
\end{Lemma}
{\Proof}
First note that $A$ cannot have a zero of infinite order at $r=\rho$.
Indeed, were this not true, we could write $(r-\rho)^2 > A(r)$ for $r$
near $\rho$, so that $A^{-1/2}(r) > (r-\rho)^{-1}$, thereby violating the
integrability of $A^{-1/2}$. Thus there exists an $s>0$ such that
\begin{equation}
g(r) \;\equiv\; \frac{A(r)}{(r-\rho)^s} \;\not \rightarrow \; 0
\spc {\mbox{as $r \searrow \rho$}}, \label{3.12}
\end{equation}
but for any $\varepsilon$, $s>\varepsilon>0$,
\begin{equation}
(r-\rho)^\varepsilon \:g(r) \;=\; \frac{A(r)}{(r-\rho)^{s-\varepsilon}}
\;\rightarrow\; 0 \spc {\mbox{as $r \searrow \rho$}}. \label{3.13}
\end{equation}
According to (\ref{3.12}), there is a positive constant $\eta$ such that for
all $r$, $\rho<r<\rho+\eta$,
\begin{equation}
\frac{1}{\sqrt{A(r)}} \;=\; (r-\rho)^{-\frac{s}{2}} \:\frac{1}{\sqrt{g(r)}}
\;\;\; ;
\label{3.14}
\end{equation}
since $A>0$, we see that $g^{-\frac{1}{2}}>0$. Pick any $\bar{r}$ in 
the interval $(\rho, \rho+\eta)$. Setting $\delta=\bar{r}-\rho$,
(\ref{3.14}) yields that
\begin{equation}
\frac{1}{\sqrt{A(\bar{r})}} \;=\; \delta^{-\frac{s}{2}}
\:\frac{1}{\sqrt{g(\bar{r})}} \;\;\;.
\label{3.15}
\end{equation}
Since $A \searrow 0$ monotonically as $r \searrow \rho$, we have $A^{-1/2}$
tending monotonically to infinity so
\begin{equation}
\frac{1}{\sqrt{A(r)}} \;>\; \delta^{-\frac{s}{2}} \:\frac{1}{\sqrt{g(\bar{r})}}
\;\;\;,\spc \rho<r<\bar{r} \;\;\;.
\label{3.16}
\end{equation}
If we now integrate (\ref{3.10}) from $\rho$ to $\bar{r}$, and note that
$(A w')(\rho)=0$, we obtain
\begin{eqnarray}
(A w')(\bar{r}) &\geq& d\:(\bar{r}-\rho) \:+\: c \:\delta^{-\frac{s}{2}}
\: \frac{1}{\sqrt{g(\bar{r})}} \:(\bar{r}-\rho) \nonumber \\
&=& d \:\delta \:+\: \frac{c}{g(\bar{r})} \:\delta^{1-\frac{s}{2}} \;\;\;,
\label{3.17}
\end{eqnarray}
where we have used (\ref{3.16}). On the other hand, writing
$B(\bar{r}):=(\sqrt{A} \:w')(\bar{r})$, we see that $B$ is uniformly
bounded near the horizon because of (\ref{2.12}). Then (\ref{3.15}) gives
\[ (A w')(\bar{r}) \;=\; \sqrt{A(\bar{r})} \:(\sqrt{A} \:w')(\bar{r})
\;=\; B \: \sqrt{g(\bar{r})} \:\delta^{\frac{s}{2}} \;\;\;. \]
Using this together with (\ref{3.17}), we get for sufficiently 
small $\delta$, and $\varepsilon>0$,
\begin{eqnarray}
B &\geq& \frac{d \:\delta^{1-\frac{s}{2}+\frac{\varepsilon}{2}}}
{\sqrt{\delta^\varepsilon g(\bar{r})}}
\:+\: \frac{c \:\delta^{1-s+\varepsilon}}{\delta^\varepsilon g(\bar{r})}
\;=\; \frac{\delta^{1-s+\varepsilon} \left[ d \:\sqrt{\delta^\varepsilon
g(\bar{r})} \:\delta^{\frac{s-\varepsilon}{2}} + c \right]}
{\delta^\varepsilon g(\bar{r})} \nonumber \\
&\geq& \frac{\delta^{1-s+\varepsilon} \left[-\frac{c}{2} + c \right]}
{\delta^\varepsilon g(\bar{r})}\spc {\mbox{if $\delta$ is small,}} \nonumber \\
&\geq& \delta^{1-s+\varepsilon} \:\frac{c}{2} \;\;\;,
\label{3.18}
\end{eqnarray}
so that $1-s+\varepsilon \geq 0$, and thus
\[ 1 \;\geq\; s \;\;\;. \]

To obtain the reverse inequality, let $0<\varepsilon<1$, and note that
\[ \frac{A(r)}{(r-\rho)^\varepsilon} \;=\; \frac{A(r)}{(r-\rho)} \:
(r-\rho)^{1-\varepsilon} \;\rightarrow\; 0 \]
as $r \searrow \rho$ because (\ref{2.3}) (with $\omega=0$) shows that $A'$
is bounded at $\rho$; thus $s \geq 1$.

We know $A(r) = (r-\rho)\: g(r)$ and $\lim_{r \searrow \rho} 
(r-\rho)^{-1} \:A(r)$ exists. Thus $\lim_{r \searrow \rho} g(r)$ 
exists and is finite. Since (\ref{3.12}) shows that
$\limsup_{r \searrow \rho} g(r) > 0$, we see that $\lim_{r \searrow \rho} g(r)
> 0$.
\QED

From (\ref{3.11}), we have
\[ A(\rho) \;=\; 0 \;\;\;,\spc A'(\rho) \;>\; 0 \;\;\;, \]
and $A'(\rho)$ is finite. 
Our Einstein metric thus has the same qualitative features as the
Schwarzschild metric.  Hence the metric singularity can be removed via a
Kruskal transformation \cite{1}.  In these Kruskal coordinates, the
Yang-Mills potential is continuous and bounded (as is easily verified).
As a consequence, the arguments from \cite{7} go through and show that the spinors
must vanish identically outside the horizon.
For this, one must notice that continuous zero order terms in the Dirac 
operator are irrelevant for the derivation of the matching conditions 
in \cite[Section 2.4]{7}. Thus the matching conditions \cite[(2.31), (2,34)]{7}
are valid without changes in the presence of our YM field.
Using conservation of the (electromagnetic) Dirac current and its 
positivity in time-like directions, the arguments of \cite[Section 4]{7}
all carry over. Thus the proof of Theorem 3.1 in the case that
$A^{- \frac 1 2}$ is integrable near the event horizon is considered 
complete.  We now turn to \\[1em]
\underline{Case 2}.  $A(r)^{-\frac 1 2}$ is not integrable near $r=\rho$.
\\[.5em]
We break the proof up into two sub-cases:
\[	{\rm (i)} \quad \omega \;\not=\; 0 \;\;\;, \spc {\rm (ii)}
\quad \omega \;=\; 0 \;\;\;.  \]
Suppose first that we are in Case (i), $\om \not= 0$.
We begin with the following lemma.
\begin{Lemma}
The function
$\left[ \left| \left( \frac m {\om T} \right)^\prime \right| + 
\left| \left( \frac w {r \om T} \right)^\prime \right| \right]$
{\it is integrable near the horizon} $r = \rho$.
\end{Lemma}
{\Proof}
We first consider $|(\frac 1 T )'|$.  To this end,
we write $\frac 1 T = \frac{\sqrt{A}} {\sqrt{A} T} $ and compute
\[ \left| \left( \frac 1 T \right)' \right| \;=\; \frac{|( \sqrt{A} T)
(\sqrt{A} )' - \sqrt{A} \:(\sqrt{A} T)'|} {AT^2}
\;\leq\; \left| \frac{( \sqrt{A} T) (\sqrt{A} )'}{A T^2} \right| +
\left| \frac{\sqrt{A} \:(\sqrt{A} T)'}{A T^2} \right| \;\;\;. \]
Now according to Assumption (I), near the event horizon $r = \rho, \
AT^2$ is bounded away from zero and $|(\sqrt{A} T)'|$ is bounded, and
thus integrable.  Moreover, since $A$ is increasing near $r = \rho$, we
see that $|\sqrt{A} T \:(\sqrt{A} )'| = (\sqrt{A} T)\:
(\sqrt{A} )'$, and since $A'>0$, this term too is integrable near $r=\rho$.
Thus $|(\frac 1 T )'|$ is integrable near $r=\rho$.

We handle $|( \frac w {rT} )'|$ similarly; namely, write
\[	\frac w {rT} \;=\; \frac {\sqrt{A} w}{r (\sqrt{A} T)}\;\;\;.	\]
Then
\begin{eqnarray*}
\left| \left( \frac w {r T} \right) ' \right| &=& \frac {|(r \sqrt{A}
T)(\sqrt{A} w)' - (\sqrt{A} w)(r \sqrt{A} T)'|} {r^2 AT^2} \\ 
&=& \frac {|r(\sqrt{A} T)[\sqrt{A} w' + (\sqrt{A})'w] - (\sqrt{A} w)
[r(\sqrt{A} T)' + \sqrt{A} T]|} {r^2 AT^2} \;\;\; ,
\end{eqnarray*}
and since $AT^2$, $w$ and $A w'^2$ are bounded and $A'>0$
near $r=\rho$, we see, as above, that $|( \frac w {rT} )'|$ is integrable
near $r = \rho$; this proves the lemma.
\QED

\begin{Prp} Assume that $\om \not= 0$; then there
exist constants $c_1 > 0, \ \ve_1 > 0$ such that
\begin{equation}
c_1 \;\le\; (\alpha^2 + \beta^2)(r) \;\le\; {1 \over c_1}
\;\;\;, \spc \rho < r < \rho + \ve_1. \label{3.20}
\end{equation}
\end{Prp}
{\Proof}
We rewrite the Dirac equations (\ref{2.1}) and (\ref{2.2}) in matrix form,
\begin{equation}
\sqrt{A} \ \Phi' = \left(
\begin{array}{cc}
\displaystyle \frac w r & -(m + \om T) \\[.5em]
-m+\om T & - \displaystyle \frac w r \end{array} \right) \Phi \;\;\;,
\label{eq:dirac}
\end{equation}
where $\Phi = (\alpha, \beta)^T$. Furthermore, let
\[  a \;=\; \om T \;\;\;, \spc b \;=\; \frac {w}{r} \;\;\;,\spc
c \;=\; -\frac{m}{\sqrt{A}} \;\;\;, \]
and define the matrix $B(r)$ by
\[ B(r) = \left( \begin{array}{cc}
\displaystyle 1 + \frac c a & \displaystyle - \frac b a \\[1em]
\displaystyle - \frac b a & \displaystyle 1 - \frac c a 
\end{array} \right) \;=\; \left( \begin{array}{cc}
\displaystyle 1 - \frac m{\om T} & \displaystyle - \frac w {r\om T} \\[1em]
\displaystyle - \frac w {r\om T} & \displaystyle 1 + \frac m{\om T}
\end{array} \right) \;\;\;. \]
Our proof is based on an extension of \cite[Lemma 5.1]{7}.
In this earlier paper we used the fact that $\frac b a$ and
$\frac c a$ are monotone near the horizon.  In the case considered
here, we do not have these hypotheses, and we must therefore work harder.

Since $T(r) \ra \infty$ as $r \searrow \rho$ and $w(r)$ is bounded
(by Hypothesis (II)), we see that both $\frac m {\om T}$ and $\frac w
{r\om T}$ tend to zero as $r \searrow \rho$.  Thus $B(r)$ is close to
the identity matrix when $r$ is near $\rho$.
  If we define $F(r)$ by
\[	F(r) = \bra \Phi(r), \ B(r) \:\Phi(r) \ket	\]
(where $\bra , \ket$ denotes the usual Euclidean inner product), then we can
find constants $c > 0$ and $\ve_1 > 0$ such that
\begin{equation}
\frac 1 c \:|\Phi(r)|^2 \;\le\; F(r) \;\le\; c \:|\Phi(r)|^2
\;\;\;, \spc \rho < r < \rho + \ve_1 . \label{3.22}
\end{equation}
Furthermore, an easy calculation using the Dirac equation (\ref{eq:dirac})
yields that
\[	F'(r) = \bra \Phi(r), \ B'(r) \:\Phi(r) \ket \;\;\;,	\]
and thus if $\rho < r < \rho + \ve_1$,
\[
-\sqrt{2} \left[ \left| \left( \frac m {\om T} \right)^\prime \right| + 
\left| \left( \frac w {r \om T} \right)^\prime \right| \right] |\Phi|^2
\;\le\; F'(r) \;\le\; 
\sqrt{2} \left[ \left| \left( \frac m {\om T} \right)^\prime \right|  + 
\left| \left( \frac w {r \om T} \right)^\prime \right| \right] |\Phi|^2
\;\;\;, \]
because the sup-norm of the matrix $B '$ is bounded by
\[ |B'| \;\le\; \sqrt{2} \left[ \left| \left( \frac m {\om T} \right)^\prime
\right| + 
\left| \left( \frac w {r \om T} \right)^\prime \right| \right] \;\;\;. \]
Thus from (\ref{3.22}), we obtain
\begin{equation}
-c \left[ \left| \left( \frac m {\om T} \right)^\prime \right| + 
\left| \left( \frac w {r \om T} \right)^\prime \right| \right] F(r) \;\le\;
\frac{F'(r)}{\sqrt{2}} \;\le\;
c \left[ \left| \left( \frac m {\om T} \right)^\prime \right| + 
\left| \left( \frac w {r \om T} \right)^\prime \right| \right] F(r)
\label{3.23}
\end{equation}
on $\rho < r < \rho + \ve_1$.
Dividing this inequality by $F(r)$ and applying Lemma~3.6, we obtain
\begin{equation}
-a(r) \;\le\; \frac{F'(r)}{F(r)} \;\le\; a(r) \;\;\;,
\label{3.24}
\end{equation}
where $a(r)$ is integrable near the event horizon $r=\rho$.  If we
integrate (\ref{3.24}) from $\rho$ to $\rho + \ve_1$, we see that $\log F(r)$
is bounded from above and below near $r=\rho$.  Upon exponentiating and
substituting into (\ref{3.22}), we see (as in the proof of Lemma 3.2)
that (\ref{3.20}) holds. \QED

We return now to the proof of Theorem 3.1 in the case $A^{- \frac 12}$
is not integrable near $r=\rho$, and $\om \not= 0$.  To this end, we
consider the differential equation for $(AT^2)$, (\ref{3.4}),
and observe that as $r \searrow \rho$ the right-hand-side tends to
$-\infty$, while the left-hand side is bounded.  This contradiction
implies that $\om \not= 0$ cannot hold.

The remaining case in the proof of Theorem 3.1 is when we assume $A^{- \frac
12}$ is not integrable near the event horizon, and $\om = 0$. In this
case, we note that Lemma 3.4 holds (cf. the remark after the statement
of this lemma).  As in Case 1 above, we find that (\ref{3.10}) holds near
$r=\rho$. Integrating this inequality from $r>\rho$ to $\rho + \ve$, we see
that for $r$ near $\rho$, the left-hand side is bounded while the
right-hand side can be made arbitrarily large.
This contradiction completes the proof of Theorem 3.1.

\begin{footnotesize}
\begin{tabular}{lclcl}
\\
Felix Finster & $\;$ & Joel Smoller & $\;$ & Shing-Tung Yau \\
Max Planck Institute MIS && Mathematics Department && Mathematics 
Department \\
Inselstr.\ 22-26 && The University of Michigan && Harvard University \\
04103 Leipzig, Germany && Ann Arbor, MI 48109, USA
&& Cambridge, MA 02138, USA \\
{\tt{Felix.Finster@mis.mpg.de}} && {\tt{smoller@umich.edu}}
&& {\tt{yau@math.harvard.edu}}
\end{tabular}
\end{footnotesize}

\end{document}